\documentstyle[psfig]{elsart}

\begin{document}
\begin{frontmatter}
\title{Quantum information theory of entanglement\\
and measurement\thanksref{talk}}
\thanks[talk]{This research was presented at the 4th Workshop
on Physics and Computation, Boston, November 1996.}
\author[NC]{Nicolas J. Cerf} and
\author[CA]{Chris Adami}
\address[NC]{Kellogg Radiation Laboratory, California Institute of 
Technology,\\ Pasadena, CA 91125, USA. E-mail: {\tt cerf@krl.caltech.edu}}
\address[CA]{Computation and Neural Systems,
California Institute of Technology,\\ Pasadena,
CA 91125, USA. E-mail: {\tt adami@krl.caltech.edu}}

\date{15 January 1997}

\begin{abstract}
We present a quantum information theory that allows for a
consistent description of entanglement.  It parallels classical
(Shannon) information theory but is based entirely on density matrices
(rather than probability distributions) for the description
of quantum ensembles.  We find
that quantum conditional entropies can be negative for entangled
systems, which leads to a violation of well-known bounds in Shannon
information theory. Such a unified information-theoretic description of
classical correlation and quantum entanglement
clarifies the link between them: the latter can be viewed as
``super-correlation'' which can induce classical correlation
when considering a tripartite or larger system.
Furthermore, negative entropy and the associated
clarification of entanglement paves the way to a natural 
information-theoretic description of the measurement process. This model,
while unitary and causal, implies the well-known probabilistic results
of conventional quantum mechanics. It also results in a simple
interpretation of the Kholevo theorem limiting the accessible information
in a quantum measurement.
\end{abstract}


\end{frontmatter}

\def\be{\begin{equation}}
\def\ee{\end{equation}}
\def\tr{{\rm Tr}}

\section{Introduction}

The recent vigorous activity in the fields of quantum information processing 
(quantum computation) and quantum communication (quantum cryptography,
teleportation, and superdense coding) has necessitated a better understanding
of the relationship between classical and quantum 
variables (see, {\it e.g.},
\cite{bib_bennett,bib_divincenzo,bib_ekertjo,bib_lloyd}).
In classical physics, information processing and communication
is best described by Shannon information theory~\cite{bib_shannon},
which succinctly associates {\em information} with randomness
{\em shared} by two physical ensembles.
Quantum information theory on the other hand is 
concerned with quantum bits (qubits) rather than bits,
and the former obey quantum laws quite
different from the classical physics of bits that 
we are used to~\cite{bib_quantumnoiseless}. 
Most importantly, qubits can exist in quantum {\em superpositions}, a notion 
essentially foreign to classical mechanics, or even classical thinking. 
To accommodate the relative phases in quantum superpositions, quantum 
information theory must be based on mathematical constructions which reflect
these: the quantum mechanical density matrices. The central object of 
information theory, the entropy, has been introduced in 
quantum mechanics by von Neumann~\cite{vonneumann} 
\be
S(\rho) = -\tr\,\rho\log\rho\;. \label{vnentropy}
\ee
Its relationship to the Boltzmann-Gibbs-Shannon entropy
\be
H(p) = - \sum_i p_i\log p_i \label{bgsentropy}
\ee
becomes obvious when considering the von Neumann entropy of a mixture
of orthogonal states. In this 
case, the density matrix $\rho$ in (\ref{vnentropy}) contains classical 
probabilities $p_i$ on its diagonal, and $S(\rho)=H(p)$. In 
general, however, quantum mechanical density matrices have off-diagonal terms,
which, for pure states, reflect the relative
quantum phases in superpositions. 
\par

In classical statistical physics, the concept of conditional and mutual
probabilities has given rise to the definition of conditional and mutual
entropies. These can be used to elegantly describe the trade-off between 
entropy and information in measurement, as well as the characteristics of a 
transmission channel. For example, for two statistical ensembles $A$ and $B$, 
the measurement of (variables of) $A$ by $B$
is expressed by the equation for the entropies
\be
H(A) = H(A|B) + H(A{\rm:}B)\;. \label{conserv}  
\ee
Here, $H(A|B)$ is the entropy of $A$ after having measured those pieces
that become correlated in $B$ [thereby apparently reducing
$H(A)$ to $H(A|B)$], while
$H(A{\rm:}B)$ is the {\em information}
gained about $A$ via the measurement of $B$.
As is well-known, $H(A|B)$ and $H(A{\rm:}B)$ compensate each other such that 
$H(A)$ is unchanged, ensuring that the second law of
thermodynamics is not violated in a measurement 
in spite of the decrease of $H(A|B)$~\cite{bib_landauer}.
Mathematically, $H(A|B)$ is a {\em conditional} entropy,
and is defined using the conditional probability $p_{i|j}$ and the 
joint probability $p_{ij}$ describing random variables from ensembles $A$
and $B$:
\be    \label{classcond}
H(A|B) = -\sum_{ij}p_{ij}\log p_{i|j}\;.
\ee
The information or mutual entropy (or correlation entropy)
$H(A{\rm:}B)$, on the other hand
is defined via the mutual probability $p_{i:j} = p_i\,p_j/p_{ij}$ as 
\be
H(A{\rm:}B) = -\sum_{ij} p_{ij} \log p_{i:j}\;. 
\ee
Simple relations such as $p_{ij} = p_{i|j}\,p_j$ imply equations such as
(\ref{conserv}) and all the other usual relations of classical information
theory. Curiously, a quantum information theory paralleling these
constructions has never been attempted. Rather, a ``hybrid'' procedure
is used in which quantum probabilities are inserted in the classical formulae
of Shannon theory, thereby losing the quantum phase crucial to density
matrices (see, {\it e.g.},~\cite{bib_zurek}). 
Below, in Section~2, 
we show that a consistent quantum information theory can be
developed that parallels the construction outlined above, while based
entirely on matrices~\cite{bib_neginfo}. This formalism allows
for a proper {\em information-theoretic} description of quantum entanglement,
unified with the standard description of classical correlations,
as shown in Section~3. As a result, most of the classical concepts
involving entropies for composite systems in Shannon theory can be extended
to the quantum regime, and this provides a simple intuitive framework
for dealing with quantum entropies.
In the fourth section, we analyze quantum
measurement in this information-theoretic language and
point out how this picture leads to a 
unitary and causal view of quantum measurement devoid of wave function 
collapse~\cite{bib_measure}. In Section~5, 
we analyze Bell-type measurements in terms of information,
as an application of this model. 
In Section~6, we conclude by considering a simple quantum information-theoretic
derivation of the Kholevo theorem (which limits the amount of information
that can be extracted in a measurement).
\par

\section{Quantum information theory}   

Let us consider the information-theoretic description of a
bipartite quantum system $AB$. A straightforward quantum generalization
of Eq.~(\ref{classcond}) suggests the definition
\be  \label{eq_defcond}
S(A|B)= - \tr_{AB} [ \rho_{AB} \log \rho_{A|B} ]
\ee
for the quantum conditional entropy. In order for
such an expression to hold, we need to define the concept of
a ``conditional'' density matrix,
\be  \label{eq_condmat}
\rho_{A|B} = \lim_{n\to\infty}
\left[ \rho_{AB}^{1/n} ({\bf 1}_A \otimes \rho_B)^{-1/n}
             \right]^n  \;,
\ee
which is the analogue of the conditional probability $p_{i|j}$. Here, 
${\bf 1}_A$ is the unit matrix in the Hilbert space for $A$,  $\otimes$
stands for the tensor product in the joint Hilbert space, and
\be
\rho_B= \tr_A [\rho_{AB}]
\ee
denotes a ``marginal'' or reduced density matrix, analogous to the marginal
probability $p_j=\sum_i p_{ij}$. The symmetrized product involving the infinite
limit in the definition of the conditional density matrix (\ref{eq_condmat})
is a technical requirement due to the fact that joint and marginal
density matrices do not commute in general.
This definition for $\rho_{A|B}$ implies that the standard relation
\be
S(A|B)=S(AB)-S(B)
\ee
holds for the quantum entropies and that $S(A|B)$ is invariant under
any unitary transformation of the product form $U_A \otimes U_B$.
More precisely, the conditional density matrix $\rho_{A|B}$ as
defined by Eq.~(\ref{eq_condmat}) is a {\em positive} Hermitian operator 
in the joint Hilbert space, whose spectrum is invariant
under $U_A \otimes U_B$.
However, in spite of the apparent similarity between
the quantum definition for $S(A|B)$ and the standard classical one
for $H(A|B)$, dealing with matrices (rather than scalars)
opens up a quantum realm for information theory exceeding
the classical one.
The crucial point is that, while $p_{i|j}$ is a probability distribution
in $i$ (i.e., $0\le p_{i|j} \le 1$), its quantum analogue $\rho_{A|B}$ 
is {\em not} a density matrix: while Hermitian and positive,
it can have eigenvalues {\em larger} than one,
and, consequently, the associated conditional entropy $S(A|B)$ can
be {\em negative}. Only such a {\em matrix-based} formalism consistently
accounts for the well-known non-monotonicity of quantum entropies
(see, {\it e.g.}, \cite{bib_wehrl}). In other words, $S(A|B)<0$
means that it is acceptable, in quantum information theory, 
to have $S(AB) < S(B)$, {\it i.e.}, the entropy of the entire system $AB$
can be smaller than the entropy of one of its subparts $B$,
a situation which is of course forbidden in classical information theory.
This happens for example in the case of quantum {\em entanglement} between 
$A$ and $B$, as will be illustrated below~\cite{bib_neginfo}.
\par

The ``non-classical'' spectrum of the conditional density matrix $\rho_{A|B}$
is related to the question of the separability of the mixed
state $\rho_{AB}$. First, the concavity of $S(A|B)$, a property related
to strong subadditivity of quantum entropies~\cite{bib_wehrl}, implies
that any separable state
\be 
\rho_{AB} = \sum_k w_k\; \rho_A^{(k)} \otimes \rho_B^{(k)}
\qquad \qquad({\rm with~} \sum_k w_k =1)
\ee
is associated with a non-negative $S(A|B)$. (Note that the converse is not
true.) Indeed, each product component  $\rho_A^{(k)} \otimes \rho_B^{(k)}$
of a separable state is associated with the conditional density matrix
\be
\rho_{A|B}^{(k)} = \rho_A^{(k)} \otimes {\bf 1}_B
\ee
so that we have
\be
S(A|B) \ge \sum_k w_k S(\rho_A^{(k)}) \ge 0   \;.
\ee
This shows that the non-negativity of conditional entropies
is a {\em necessary} condition for separability. This condition
can be shown to be equivalent to the non-violation of entropic 
Bell inequalities~\cite{bib_bell}.
Secondly, it is easy to check
from Eq.~(\ref{eq_defcond}) that, if $S(A|B)$ is negative, $\rho_{A|B}$
must admit at least one ``non-classical'' eigenvalue
({\it i.e.}, an eigenvalue exceeding one), while the converse 
again does not hold.
This results from the fact that $\tr(\rho \sigma) \ge 0$ if $\rho$ and
$\sigma$ are positive Hermitian matrices.
This suggests the conjecture that a strong necessary condition
for separability is that {\it all} the eigenvalues
of $\rho_{A|B}$ are ``classical'' ($\le 1$). When applied to the
case of a Werner state (an impure singlet state), this separability
condition turns out to be necessary (and sufficient for a $2\times 2$
Hilbert space)~\cite{bib_neginfo}, since it reduces exactly
to the condition derived in Ref.~\cite{bib_peresprl} by
considering the positivity of the partial transpose of $\rho_{AB}$.
When applied to a randomly generated mixture of product states,
this condition is always fulfilled ({\it i.e.}, all the eigenvalues
of $\rho_{A|B}$ and $\rho_{B|A}$ are $\le 1$). This opens the possibility
that it could be a stronger necessary (or perhaps even
sufficient) condition
for separability in a Hilbert space of arbitrary dimensions.
Further work will be devoted to this question.
\par

Similarly to what we have done for the conditional entropy,
the quantum analogue of the mutual entropy can be
constructed by defining a ``mutual'' density matrix
\be   \label{eq_mutmat}
\rho_{A:B} = \lim_{n\to \infty} 
\left[  (\rho_A \otimes \rho_B)^{1/n}\rho_{AB}^{-1/n}
             \right]^n \;,
\ee
the analogue of the mutual probability $p_{i:j}$. As previously, this
definition implies the standard relation 
\be  \label{eq_quantummutual}
S(A{\rm:}B)=S(A)-S(A|B)=S(A)+S(B)-S(AB)
\ee
between the quantum entropies. This definition extends
the classical notion of {\em mutual information} or correlation entropy
$H(A{\rm:}B)$ to the quantum notion of {\em mutual entanglement}
$S(A{\rm:}B)$. Note that all the above quantum definitions reduce to the
classical ones for a diagonal $\rho_{AB}$, which suggests that
Eqs. (\ref{eq_condmat}) and (\ref{eq_mutmat}) are reasonable assumptions.
(It is possible that other definitions of $\rho_{A|B}$ and $\rho_{A:B}$
can be proposed, but we believe this choice is simplest.)
The proposed matrix-based information theory therefore includes
Shannon theory as a special case, while it describes quantum entanglement
as well. Since the definition of mutual entanglement $S(A{\rm:}B)$
covers classical correlations also, $S(A{\rm:}B)$ must be considered
as a general measure of correlations {\em and} ``super-correlations''
in information theory, which applies to pure as well as
mixed states. It is worth noticing that this does {\it not} mean that
the mutual entanglement characterizes the purely quantum correlation
between $A$ and $B$ (that part which can be purified to singlet states);
rather $S(A{\rm:}B)$ does not separate correlation and entanglement,
it is a measure of both. In this sense, $S(A{\rm:}B)$ differs
from various definitions of the entropy of entanglement which can be found
in the literature~\cite{bib_mixed}. Finally, 
we show in Ref.~\cite{bib_channel} that,
besides being the proper quantum counterpart of correlation entropy,
$S(A{\rm:}B)$ also turns out to be a basic quantity in the search for
a quantum counterpart of Shannon's fundamental coding theorem
on noisy channels. Indeed, our proposed definition for the capacity
for entanglement transmission through a quantum channel
is written as the maximum achievable mutual entanglement $S(A{\rm:}B)$,
in analogy with the classical definition.
\par

\begin{figure}
\caption {(a) General entropy diagram for a quantum bipartite system $AB$.
(b) Entropy diagrams for three cases of a system of 2 qubits: (I) independent,
(II) classically correlated, (III) quantum entangled.}
\label{figAB}
\vskip 0.25cm
\centerline{\psfig{figure=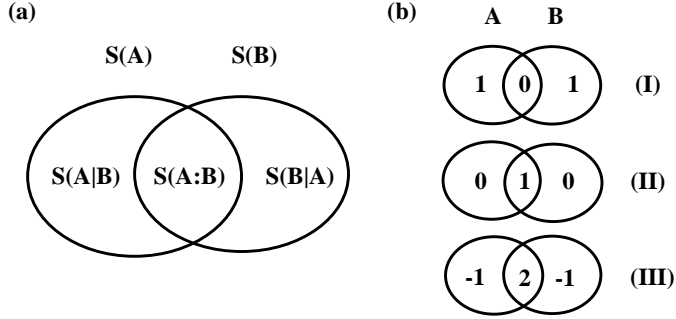,width=3.50in,angle=0}}
\end{figure}

As we shall see in the next section,
our quantum matrix-based formalism can be successfully applied to the
quantum entanglement of more than two systems by extending the
various classical entropies that are defined in 
the Shannon information-theoretic
treatment of a multipartite system. This accounts for
example for the creation of classical correlation through quantum
entanglement in a tripartite (or larger) system. Also,
the quantum analogue of all the fundamental relations
between classical entropies (such as
the chain rules for entropies and mutual entropies) holds in quantum
information theory and have the same intuitive interpretation,
and we make extensive use of it in~\cite{bib_channel,bib_measure,bib_bell}.
Let us close this section by suggesting a simple diagrammatic way
of representing quantum entropies which provides intuitive insight
into this information-theoretic description of entanglement.
In the case of a bipartite system, the relations between $S(A)$, $S(B)$,
$S(AB)$, $S(A|B)$, $S(B|A)$, and $S(A{\rm:}B)$ are conveniently summarized
by a Venn-like entropy diagram, as shown in Fig.~\ref{figAB}a.
The important difference
between classical and quantum entropy diagrams is that the basic
inequalities relating the entropies are ``weaker'' in the quantum case,
allowing for negative conditional entropies and ``excessive'' mutual 
entropies~\cite{bib_neginfo}.
For example, the upper bound for the mutual entropy (which is directly
related to the channel capacity) is 
\be
H(A{\rm:}B) \le \min[H(A),H(B)]
\ee
in classical information theory, as a consequence of
the inequality $H(AB) \ge \max[H(A),H(B)]$, while it is
\be
S(A{\rm:}B) \le  2 \min[S(A),S(B)]
\ee
in quantum information theory, as a result of the Araki-Lieb
inequality~\cite{bib_wehrl} $S(AB) \ge |S(A)-S(B)|$.
This means that a quantum channel has a capacity for entanglement transmission
that can reach twice the classical upper
bound~\cite{bib_channel,bib_neginfo}; this is
apparent for instance in the superdense coding scheme~\cite{bib_supercod}.
\par

\section{Correlation versus entanglement and multipartite systems}

We show in Fig.~\ref{figAB}b the entropy diagram corresponding to three
limiting cases of a bipartite  system of two dichotomic variables
(e.g., 2 qubits): independent variables (case I),
classically correlated variables (case II), and quantum entangled
variables (case III). In all three cases, each subsystem taken separately
is in a mixed state of entropy $S(A)=S(B)=1~$bit.
Cases I and II correspond to classical situations
(which can of course be described in our matrix-based formalism
as well, using diagonal matrices),
while case III is a purely quantum situation which violates
the bounds of classical information theory~\cite{bib_neginfo}. 
Let us focus on case III, since cases I and II are standard.
This case corresponds to an EPR pair\footnote{Although we use the term
``EPR state'' for the wave-function~(\ref{eq_EPR}),
this state is in fact one of the {\em Bell} states, which are a generalization
of the EPR singlet state.}, characterized by the pure state
\be  \label{eq_EPR}
| \psi_{AB} \rangle = {1 \over \sqrt{2}} (|00\rangle + |11\rangle) \;,
\ee
and, accordingly, it is associated with a vanishing total entropy
$S(AB)=0$.
Using the density matrix of the joint system
$\rho_{AB}=|\psi_{AB}\rangle \langle \psi_{AB}|$,
we see that subpart $A$ (or $B$) has the marginal density matrix
\be
\rho_A = \tr_B [\rho_{AB}] =
       {1\over 2} (|0\rangle \langle0| +|1\rangle \langle1|)  \;,
\ee
and is therefore in a mixed state of positive entropy. This purely
quantum situation corresponds to the unusual entropy diagram (--1,2,--1)
shown in Fig.~\ref{figAB}b. That the EPR situation cannot be described
classically is immediately apparent when considering the associated
density matrices.
The joint and the marginal density matrices
can be written in basis $\{ 00,01,10,11 \}$ as
\be   \label{eq_jointmat}
\rho_{AB}= \left(
\begin{array}{cccc}
1/2 & 0 & 0 & 1/2 \\
0 & 0 & 0 & 0 \\ 
0 & 0 & 0 & 0 \\ 
1/2 & 0 & 0 & 1/2
\end{array}  \right) \;, \qquad
\rho_{A}= \rho_B = \left(
\begin{array}{cc}
1/2 & 0  \\ 
0 & 1/2
\end{array}  \right) \;.
\ee 
so that we obtain for the conditional density matrix\footnote{Note that for
EPR pairs, joint and marginal density matrices commute, simplifying
definitions (\ref{eq_condmat}) and (\ref{eq_mutmat}).}
\be   \label{eq_condmatex}
\rho_{A|B}= \rho_{AB} ({\bf 1}_A \otimes \rho_B)^{-1} = \left(
\begin{array}{cccc}
1 & 0 & 0 & 1 \\
0 & 0 & 0 & 0 \\ 
0 & 0 & 0 & 0 \\ 
1 & 0 & 0 & 1
\end{array}  \right) \;.
\ee 
Plugging (\ref{eq_jointmat}) and (\ref{eq_condmatex}) into
definition (\ref{eq_defcond}) immediately yields
$S(A|B)=-1$, which results in
\begin{equation}
S(AB)=S(A)+S(B|A)=1-1=0
\end{equation}
as expected. This is a direct consequence of the fact
that $\rho_{A|B}$ has one ``non-classical'' ($>1$) eigenvalue, 2.
It is thus misleading to describe an EPR-pair (or any of the Bell states)
as a correlated state within Shannon information theory, since negative
conditional entropies are crucial to its description\footnote{In
Ref.~\cite{bib_neginfo},
we suggest that EPR pairs are better understood in terms of a qubit-antiqubit
pair, where the qubit (antiqubit) carries plus (minus) one bit, and where
antiqubits are interpreted as qubits traveling {\em backwards} in time.}.
Still, classical {\em correlations} [with entropy diagram $(0,1,0)$]
emerge when {\em observing} an entangled EPR pair. Indeed, after measuring 
$A$, the
outcome of the measurement of $B$ is known with 100\% certainty.
The key to this discrepancy lies in the information-theoretic
description of the measurement process~\cite{bib_measure}.
Anticipating the next section, let us just mention
that the {\em observation} of quantum entangled states such as an EPR pair
gives rise to classical correlations between the two measurement {\em devices}
while keeping the entanglement between the two parties 
(particle + measurement device) unchanged,
thereby creating the confusion between entanglement and correlation.
\par

\begin{figure}
\caption {Ternary entropy Venn-diagram for a general tripartite
system $ABC$. The component entropies are defined in the text.}
\label{fig_general3}
\vskip 0.25cm
\centerline{\psfig{figure=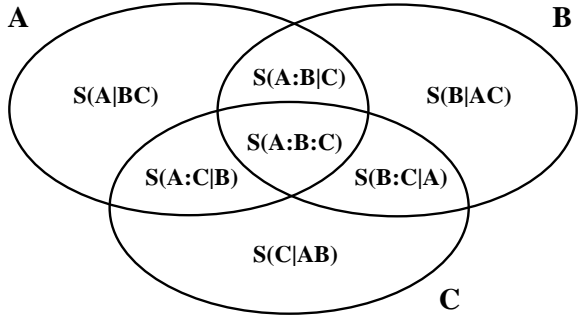,width=3.0in,angle=-90}}
\end{figure}

More generally,
the concept of negative conditional entropy turns out to be very
useful to describe {\it multipartite}
quantum systems, and it gives new insight into
the creation of classical correlations from quantum
entanglement. In the case of a tripartite system, the quantum
entropies involved can be represented by a Venn-like diagram, as shown in
Fig.~\ref{fig_general3}. The conditional entropies $S(A|BC)$,
$S(B|AC)$, and $S(C|AB)$ are a straightforward generalization
of conditional entropies in a bipartite system, {\it i.e.},
$S(A|BC)=S(ABC)-S(BC)$, etc. The entropies $S(A{\rm:}B|C)$,
$S(A{\rm:}C|B)$, and $S(B{\rm:}C|A)$ correspond to conditional mutual
entropies. They characterize the mutual entanglement between two of the
subsystems when the third is known. In perfect analogy with the classical
definition, one can write, for example,
\be
S(A{\rm:}B|C)=S(A|C)-S(A|BC)  \;.
\ee
This is a straightforward generalization of
Eq.~(\ref{eq_quantummutual}) where all the entropies are 
conditional on $C$. A trivial calculation gives also the expression
of the conditional mutual entropy in terms of total entropies
\be        \label{eq_condmutual}
S(A{\rm:}B|C)=S(AC)+S(BC)-S(C)-S(ABC)
\ee
This last expression illustrates that the conditional mutual entropies
are always non-negative as a consequence of strong subadditivity
of quantum entropies (see, {\it e.g.}, \cite{bib_wehrl}), a property
that will be useful in the following.
The entropy in the center of the diagram is
a {\it ternary} mutual entropy, defined as
\be 
S(A{\rm:}B{\rm:}C)=S(A{\rm:}B)-S(A{\rm:}B|C)
\ee
(this generalizes Eq.~(\ref{eq_quantummutual}) for a mutual
entropy rather than a total entropy). Using Eq.~(\ref{eq_condmutual}),
this can be written in a more symmetric way as
\begin{eqnarray}
S(A{\rm:}B{\rm:}C)&=&S(A)+S(B)+S(C)-S(AB)-S(AC)-S(BC) \nonumber \\
&+&S(ABC)   \;.
\end{eqnarray}
More generally, relations between entropies in a multipartite system
can be written, such as the ``chain rules'' for quantum entropies
\be  \label{eq_chainrule1}
S(A_1\cdots A_n)=S(A_1)+S(A_2|A_1)+S(A_3|A_1A_2)+\cdots
\ee
or for quantum mutual entropies
\be  \label{eq_chainrule2}
S(A_1\cdots A_n{\rm:}B)=S(A_1{\rm:}B)+S(A_2{\rm:}B|A_1)
+S(A_3{\rm:}B|A_1A_2)+\cdots
\ee

\begin{figure}
\caption {(a)~Ternary entropy diagram for an ``EPR-triplet''
or GHZ state.
(b)~Entropy diagram for subsystem $AB$ unconditional on $C$. }
\label{figABC}
\vskip 0.25cm
\centerline{\psfig{figure=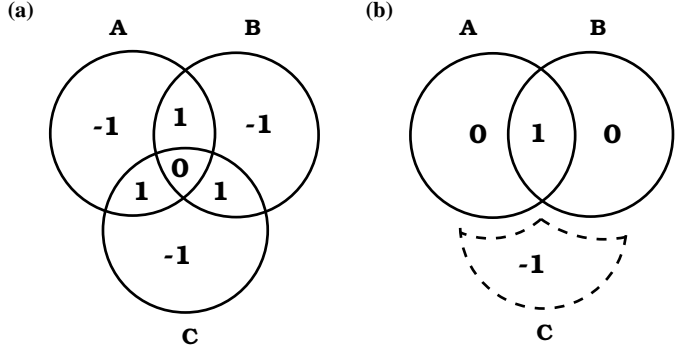,width=3.50in,angle=-90}}
\vskip -0.25cm
\end{figure}

Let us consider as an illustration a tripartite system $ABC$
in a Greenberger-Horne-Zeilinger (GHZ)~\cite{bib_GHZ}
state\footnote{The GHZ state can
also be viewed as an ``EPR-triplet'', a generalization
of an EPR-pair to three parties.},
\be
| \psi_{ABC} \rangle = {1 \over \sqrt{2}} (|000\rangle + |111\rangle)  \;.
\ee
As it is a pure (entangled) state, the total entropy is $S(ABC)=0$. 
The corresponding ternary entropy diagram of $ABC$ is shown
in Fig.~\ref{figABC}a. Note that the vanishing {\em ternary}
mutual entropy 
\be
S(A{\rm:}B{\rm:}C)=0
\ee
in the center of the diagram is generic to any entangled tripartite
system in a pure state~\cite{bib_measure}\footnote{For a multipartite system,
the mutual entropy between the $n$ parts is equal to $1+(-1)^n$.}.
Indeed, the Schmidt
decomposition of the pure state $| \psi_{ABC} \rangle$ implies
that $S(AB)=S(C)$, $S(AC)=S(B)$, and $S(BC)=S(A)$. This feature
will be important in the following section as it implies that no information
(in the sense of Shannon theory) is extracted in the measurement
of a pure state. Fig.~\ref{figABC}a shows clearly
that, grouping say $A$ and $B$ (considering them as a single entity)
results in entanglement [diagram (-1,2,-1)] between $AB$ and $C$.
On the other hand, when tracing over the degree of 
freedom associated with $C$, say, the 
resulting marginal density matrix for subsystem $AB$ is
\be
\rho_{AB} = \tr_C [\rho_{ABC}] =
       {1\over 2} (|00\rangle \langle 00| +|11\rangle \langle 11|) \;,
\ee
corresponding to a classically correlated system [diagram (0,1,0)]. As the
density matrix {\em fully} characterizes a quantum system, subsystem $AB$
(unconditional on $C$, i.e., ignoring the existence of $C$)
is in this case physically {\em indistinguishable} from a statistical ensemble
prepared with an equal number of $|00\rangle$ and $|11\rangle$ states.
Thus, $A$ and $B$ are correlated in the sense of Shannon theory
if $C$ is ignored. The ``tracing over'' operation depicted in 
Fig.~\ref{figABC}b illustrates this creation of
classical correlation from quantum entanglement. 
In short, the EPR-triplet entails quantum entanglement between any
part, e.g. $C$, and the rest of the system $AB$. 
The subsystem $AB$ {\em unconditional} on $C$ has a positive entropy
$S(AB)$ of 1~bit, and is indistinguishable from a classical
correlated mixture. On the other hand, the
entropy of $C$ conditional on $AB$, $S(C|AB)$, is negative and equal to
$-1$~bit, thereby counterbalancing $S(AB)$ to yield a vanishing combined
entropy
 \be
S(ABC)=S(AB)+S(C|AB)=0  \;,
 \ee 
as expected in view of the entanglement between $AB$ and $C$.
The above can be extended in a straightforward manner to composite 
systems, and this will be central to the measurement process.
\par

\section{Quantum measurement}

According to von Neumann~\cite{vonneumann}, a consistent description
of the measurement process must involve the interaction between
the observed quantum system and a {\em quantum} measurement device.
Such a view is in contrast with the Copenhagen interpretation of quantum 
mechanics (see, e.g.,~\cite{wheeler_zurek}) stating that the
measurement is the non-causal process of projecting the wave-function,
which results from the interaction with a {\em classical} apparatus.
A classical apparatus is defined as one where the ``pointer'' 
variables take on {\em definite} values, and which therefore cannot
reflect quantum superpositions.
For 70 years, the Copenhagen interpretation has never failed in predicting
a single experimental fact, which certainly has helped in cementing its
reputation~\cite{wheeler_zurek}.
On the other hand, if the foundations of quantum 
mechanics are believed to be solid, it cannot be denied that measurement is 
{\em not} an abstract non-causal operation acting on wave functions,
but rather a genuine interaction between two physical {\em quantum} systems:
the observed system $Q$ and the measurement device, or the ancilla $A$. 
This is the essence of the von Neumann theory of measurement. 
\par

Assume then that a quantum system is initially in state
\be   \label{eq_superposition}
|Q\rangle = \sum_i \alpha_i |a_i\rangle
\ee
expressed in the basis $\{ |a_i\rangle \}$ of eigenvectors of an
arbitrary observable (the one that we are measuring). Then, 
the von Neumann measurement is described by the {\em unitary} transformation
that evolves the initial state of the joint system $|Q,A\rangle=|Q,0\rangle$
into the state
\be   \label{eq_entangl}
|QA\rangle = \sum_i \alpha_i | a_i,i \rangle
\ee
with $\{|i\rangle\}$ denoting the eigenstates of the ancilla
($|0\rangle$ is the reference initial state of the ancilla). Such a 
transformation was 
interpreted by von Neumann as inducing {\em correlations} between the system 
$Q$ and
the ancilla $A$. Indeed, if $|Q\rangle$ is initially in one of the eigenstates
$|a_i\rangle$ (i.e., if it is {\em not} in a superposition),
the ``pointer'' in $A$ that previously pointed to zero now points
to the eigenvector $|i\rangle$ which labels outcome $i$, suggesting
that a measurement has been performed.
\par

Now, a basic problem occurs if the initial state of $Q$ is
a superposition, as in Eq.~(\ref{eq_superposition}), that is, if $Q$
is {\em not} in an eigenstate of the considered observable.
Then, according to Eq.~(\ref{eq_entangl}),
the apparatus apparently points to a {\em superposition}
of $i$'s, a fact which obviously contradicts our everyday-life
experience. In classical physics, a variable has, at any time, a definite
value that can be recorded. Experiments show that a quantum
measurement is probabilistic in nature, that is {\em one} of the possible 
outcomes
(drawn from a probability distribution) becomes {\em factual}. In other
words, a quantum superposition evolves into a mixed state.
This apparent necessity led von Neumann to introduce an {\it ad hoc},
non-unitary,
{\em second} stage of the measurement, called {\em observation}. In this
stage, the measurement is ``observed'', and a collapse occurs in order
to yield a classical result from a quantum superposition. 
The central point in the quantum information-theoretic interpretation
of the measurement problem presented below (see also Ref.~\cite{bib_measure})
is that, in general,
the state described by Eq.~(\ref{eq_entangl}) is {\em entangled}, 
not just correlated. As emphasized earlier, entangled states have an
information-theoretic description distinct from correlated states,
which provides them with very peculiar properties.
For example, it has been shown that an arbitrary quantum 
state cannot be {\em cloned}~\cite{bib_noncloning} precisely because of the 
entanglement between the system $Q$ and the ancilla $A$. If the system is in a
state belonging to a set of {\em orthogonal} states, on the other hand, 
a faithful copy of the quantum state can be obtained applying a von Neumann 
measurement. As a consequence it appears that an {\it arbitrary} state
(one which is {\em not} one of the eigenstates of the observable considered)
can not be {\em measured} without creating entanglement.
\par

Let us show that unitary evolution [such as the one giving rise to
Eq.~(\ref{eq_entangl})] can be
reconciled with the creation of randomness in the measurement process
if it is recognized that the creation of entanglement (rather than
correlation) is {\em generic} to a quantum measurement, {\em and} if this 
entanglement is properly described in quantum information theory
using the concept of negative entropy (see also Ref.~\cite{bib_measure}).
This reconciliation is brought about by a redescription of the second stage
of measurement, the observation, without involving an irreversible loss
of information to a macroscopic environment. In that respect, our
model is distinct from the environment-induced decoherence model,
one of the prevalent contemporary views of quantum measurement
(see {\it e.g.}~\cite{bib_zurekphystod}).
In order to observe the measurement, a system involving
generally a large number of degrees of freedom has to interact with
$Q$. In Eq.~(\ref{eq_entangl}), $Q$ has interacted with a single
degree of freedom of the ancilla $A$ (first stage of the measurement),
which led to an entangled state. As emphasized before, the creation of an 
entangled state does not mean that a measurement has 
been performed, since our (classical) perception of a measurement is 
intrinsically related to the existence of (classical) correlations. In order 
for classical correlations to emerge, a third degree of freedom (another 
ancilla $A'$) has to be involved. Now, iterating the von Neumann
measurement, $A'$ interacts with $AQ$ so that the resulting state
of the combined system is
\be   \label{eq_QAA}
|QAA'\rangle = \sum_i \alpha_i | a_i,i,i \rangle  \;,
\ee
where the eigenstates of $A'$ are also denoted by $|i\rangle$ for simplicity.
The state so created is pure [$S(QAA')=0$], akin to an ``EPR-triplet''
since the system has undergone only unitary transformations
from a pure initial state $|Q,0,0\rangle$.
The point is that, considering the state of the entire
ancilla $AA'$ {\em unconditionally} on system $Q$ yields a {\em mixed state}
\be \label{classcorr}
\rho_{AA'}= \tr_Q [\rho_{QAA'}]=
\sum_i |\alpha_i|^2  |i,i \rangle \langle i,i |
\ee
describing maximal correlation between $A$ and $A'$, that is
\be   \label{eq_correlationancillae}
S(A{\rm:}A')=S(A)=S(A')=S(AA') \;.
\ee 
The second stage
consists in observing {\em this} classical correlation (that
extends, in practice, to the $10^{23}$ particles which constitute 
the macroscopic measurement device). Note that a macroscopic measurement
device is not required here, since only two ancillary degrees of freedom
$A$ and $A'$ are enough to generate correlation in the tripartite entangled
system $QAA'$. The entropy diagram characterizing $QAA'$ is of the same kind
as the one depicted in Fig.~\ref{figABC}, but filled in
with a constant different from 1 in general.
Paradoxically, it is the physical state
of the {\em ancilla} which contains the outcome of the measurement, whereas
the quantum state $Q$ itself must be {\em ignored} to observe correlations.
This crucial point is easily overlooked,
since intuition dictates that performing a measurement means somehow
``observing the state of $Q$''. Rather, a measurement is constructed such
as to {\em infer} the state of $Q$ from that of the ancilla---but ignoring $Q$
itself. The correlations (in $AA'$) which emerge from the fact that a 
part ($Q$) of an entangled state ($QAA'$) is ignored give rise to the 
classical idea of a measurement. This view of the measurement process
insists only on the ``self-consistency'' of the measurement device, while
abandoning the 100\% correlation between the latter and the quantum system $Q$,
a cornerstone of decoherence models. More precisely, no information
(in the sense of Shannon theory) about $Q$ is obtained from the ancilla.
Indeed, using Eqs.~(\ref{eq_condmutual}) and (\ref{eq_correlationancillae}),
we have 
\be
S(Q{\rm:}A{\rm:}A')=S(A{\rm:}A')-S(A{\rm:}A'|Q)=0
\ee
meaning that the mutual entropy between $A$ and $A'$ (the observed correlation)
is {\it not} shared with $Q$. This is a consequence of the fact that
$Q$ is initially in a pure state. We will see in the next section
that information (Shannon mutual entropy) can only be acquired
in the situation where a {\it mixed} state is measured.
After measurement, the quantum entropy of the ancilla (unconditional on $Q$)
\be
S(AA')=H[p_i] \qquad {\rm with~}p_i=|\alpha_i|^2 
\ee
is interpreted as the ``physical'' entropy of $Q$. This happens
to be the classical entropy associated with the probability distribution 
of the random outcomes, $p_i=|\alpha_i|^2$, that is
the probabilities predicted by quantum mechanics.
Thus the unconditional entropy of the ancilla is equal to the entropy
of $Q$ predicted in ``orthodox'' quantum mechanics
(which involves the projection of the wave function).
Still, the entropy of $Q$ conditional on $AA'$ is {\em negative}, and
exactly compensates $S(AA')$ to allow for a vanishing entropy for the
joint system,
\be
S(AA')+S(Q|AA')=S(QAA')=0
\ee
This then emphasizes how measurement can be probabilistic
in nature, while at the same time being described by a unitary
process (which does {\em not} permit the evolution of pure into mixed states).
\par

The appearance of a wave-function collapse, crucial in
the physics of sequential measurements, can also be interpreted in
this information-theoretic picture.
If a second ancilla $B$ (in general, also a large
number of degrees of freedom) interacts with $Q$ in order to
measure the {\em same} observable (after a first measurement involving
ancilla $A$), the result is an ``EPR-$n$plet''
(consisting of all the degrees of freedom of $A$, $B$, and the measured 
quantum state $Q$). To simplify, let us consider two ancillary variables
$A$ and $B$ (and neglect their amplification).
Then, the final quantum state after the sequential measurement is
\be   \label{eq_QAABB}
|QAB\rangle = \sum_i \alpha_i | a_i,i,i \rangle
\ee
illustrating clearly that the state of $A$ and $B$ (unconditional
on $Q$) are {\em classically} maximally correlated just as described earlier.
This is the basic consistency
requirement for two consecutive measurements of the same variable:
we must have $S(B|A)=0$.
The standard assertion of orthodox quantum mechanics is that, after
the first measurement, the wave function of $Q$ is projected on
$|a_i\rangle$, the observed eigenstate, so that any following measurement
yields the same value $i$ without any remaining uncertainty since the
state of $Q$ is now $|a_i\rangle$. As we just showed, 
such a classical correlation between the {\em outcome} of two measurements
actually involves {\em no} collapse; rather, the vanishing remaining
uncertainty of the second measurement [reflected by the
vanishing conditional entropy $S(B|A)=0$] is due to the fact that one
considers only part of an entangled system.
\par

More interestingly, in the case where the ancilla $B$ measures another
observable, Eq. (\ref{eq_QAABB}) becomes then
\be
|QAB\rangle = \sum_{i,j} \alpha_i U_{ij} | b_j,i,j \rangle
\ee
where $\{|b_j\rangle\}$ are the eigenvectors of the second observable,
$U_{ij}=\langle b_j | a_i \rangle$, 
and $\{|j\rangle\}$ denote the eigenstates of the second ancilla $B$.
The resulting entropy diagram for $AB$ (obtained by tracing over $Q$)
gives rise to an {\em entropic} uncertainty relation~\cite{bib_measure}
\be \label{eq_uncert}
S(A)+S(B) \ge \min_i H \left[ |U_{ij}|^2 \right]_{i~{\rm fixed}}
\ee
where the right-hand side stands for the Shannon entropy of the
probability distribution resulting from the expansion of the
eigenvector $|a_i\rangle$ of the first observable into the eigenbasis
of the second observable, minimized over $i$. This reflects the fact
that the sequential measurement of two non-commuting observables
(such that $U_{ij}$ is not the identity or a permutation matrix)
must generate a non-zero entropy. Eq.~(\ref{eq_uncert}) is compatible
with the uncertainty relations found in the literature 
(see, {\it e.g.}, \cite{bib_deutsch}),
but can be shown to be stronger in intermediate
situations between compatible and
complementary (maximally incompatible) observables~\cite{bib_measure}.
\par

\section{Bell-type measurements}

In order to illustrate the information-theoretic analysis
of measurement described above, let us consider the measurement of
an EPR pair. This should also
clarify how quantum entanglement can have the appearance of classical
correlation in such an experiment.
Let us prepare a bipartite system $Q_1 Q_2$ in the EPR-entangled state
\be
|Q_1 Q_2\rangle = {1\over\sqrt{2}}
(|\uparrow\uparrow\rangle + |\downarrow\downarrow\rangle)
\ee
and separate the two members at remote locations in space. 
At each location, the system ($Q_1$ or $Q_2$) is measured by interacting
with an ancilla ($A_1$ or $A_2$), following the same procedure as
before. In brief, each system ($Q_1$ or $Q_2$) becomes entangled
with its corresponding ancilla, resulting in the entangled state
\be
|Q_1 Q_2 A_1 A_2\rangle = {1\over\sqrt{2}}
(|\uparrow\uparrow 1 1\rangle + |\downarrow\downarrow 0 0\rangle)
\ee
for the entire system. Note that an ancilla in state $|1\rangle$ means
that a spin-up has been measured, and conversely. 
(Obviously, this corresponds to the measurement of the spin-projection
along the z-axis; the measurement of different spin-components
of $Q_1$ and $Q_2$ can be considered along the same lines.)
As previously, we describe the ancilla with just one internal variable,
even though in practice it must be thought of as consisting of a large 
number of correlated ones. The important point here is that, despite the 
fact that $Q_1$ and $Q_2$ were initially in an entangled state 
[characterized by the $(-1,2,-1)$ entropy diagram],
the state of the two ancillae
unconditional on $Q_1$ and $Q_2$ is a mixed (classically correlated) state
\be
\rho_{A_1 A_2} =
       {1\over 2} (|00\rangle \langle 00| +|11\rangle \langle 11|)
\ee
Thus, the ancillae are {\em correlated}: the corresponding 
entropy diagram $(0,1,0)$ clearly shows that,
after observing $A_1$, for instance, the state
of $A_2$ can be inferred without any uncertainty,
{\it i.e.}, $S(A_2|A_1)=0$. However, this must {\it not}
be attributed to the existence of classical correlation between
$Q_1$ and $Q_2$; rather it is the act of measuring which gives rise to 
this appearance. 
\begin{figure}
\caption {Ternary entropy diagram for the measurement of an EPR
pair. (a)~$A_1$ and $A_2$ both measure the spin $z$-component $\sigma_z$.
(b)~$A_1$ measures $\sigma_z$ while $A_2$ measures $\sigma_x$. }
\label{figEPRmeas}
\vskip 0.25cm
\centerline{\psfig{figure=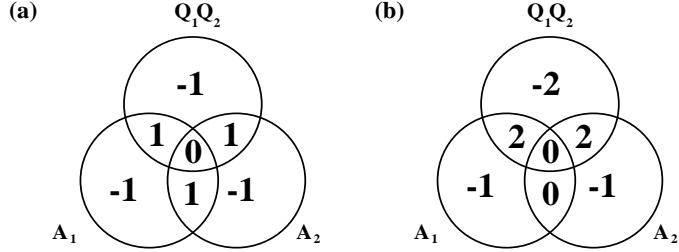,width=3.5in,angle=-90}}
\end{figure}
The entropy relations between $Q_1$, $Q_2$, $A_1$ and
$A_2$ can be summarized by an entropy diagram (Fig.~\ref{figEPRmeas}a).
This emphasizes that it is the same mechanism which is at the origin of the
coincidence between the observed spin-projection for both particles
in an EPR experiment and at the core of the consistency between sequential
measurements on a single quantum system.\footnote{We thank Zac Walton
for pointing this out to us.} In the former case,
the mechanism is well known and accepted, while in the latter case,
it is more difficult to discern and an {\it ad hoc} collapse is
therefore often wrongly invoked.
For completeness, the entropy diagram describing the situation where
the ancillae $A_1$ and $A_2$ measure {\it orthogonal} spin projections
({\it e.g.}, $\sigma_x$ and $\sigma_z$) is shown in Fig.~\ref{figEPRmeas}b.
When tracing over $Q_1$ and $Q_2$, it is obvious that the ancillae $A_1$
and $A_2$ are statistically independent, which accounts for
the fact that two apparently independent random variables 
($\sigma_x$ and $\sigma_z$) are measured. In reality,
the entire system is entangled in a particular way:
$A_1$ and $A_2$ are entangled {\it separately} with $Q_1Q_2$.
Note finally that the violation of Bell inequalities occurring
in the measurement of EPR pairs can also be analyzed from
an information-theoretic point of view, as shown in Ref.~\cite{bib_bell}.

\section{Measurement of mixed states and accessible information}

An important issue of quantum information theory is the maximum amount of
information that one can extract about a quantum system by performing
a measurement. Let us consider a simple derivation
of this quantity based on conditional and mutual quantum entropies 
and on simple relations between them. This derivation, akin to
the proof of Schumacher {\it et al.}~\cite{bib_schumacher}, relies on our
information-theoretic description of unitary measurement
and does not involve any ``environmental'' degrees of freedom
(it does not involve decoherence induced by an
environment~\cite{bib_zurekphystod}). As emphasized before,
the entropy that appears in the ancilla $A$ is ``extracted''
from the measured quantum system $Q$, whose conditional quantum entropy
therefore becomes negative.
This means that the quantum system and the ancilla
are entangled as a result of the measurement, and that the
measurement simply becomes the ``act'' of ignoring -- or tracing over --
the quantum system $Q$
which is aimed to be measured. This is in contrast with
the prevalent view of measurement, where the quantum system
and the ancilla become classically correlated because one is compelled
to ignore the numerous degrees of freedom of an uncontrollable
environment (in other words, decoherence leads to the selection of a
``preferred basis'' for the ``pointer variable''~\cite{bib_zurekphystod}).
As stressed in Section~4, the appearance of a collapse of the wave function
can be fully understood when considering subsequent
measurements {\em without} any environment; the statistics of the state
of the ancillae that interact sequentially with the quantum system $Q$
is indistinguishable from the statistics resulting
from the collapse postulate.

A striking consequence of this information-theoretic interpretation
is that, in any measurement of a {\it pure} state,
no information at all (in the sense of Shannon theory) can possibly
be extracted from the system. In other words, no information is gained
about the identity of the pure state. (This means that the
``pointer variable'' is {\it not} classically
correlated with the quantum system.) Recognizing
that a pure state has a vanishing von Neumann entropy, this
turns out to be an obvious result: there is no uncertainty about it,
so nothing can be learned from it.
This can also be understood as a consequence of the quantum
non-cloning theorem~\cite{bib_noncloning}: one cannot ``clone''
({\it i.e.}, correlate in the Shannon sense) an arbitrary state with
an ancilla, as only entanglement results from the measurement.
It is also straightforward to see, by looking at quantum entropies, that the
correlations that appear in a measurement do not concern
$Q$ {\it vs.} $A$ but rather concern all the
pieces of the (generally macroscopic) ancilla: the ancilla is
``self-consistent''\footnote{If $A_1$ and $A_2$ represent two
halves (arbitrarily chosen) of the ancilla, the ternary mutual entropy
$S(A_1{\rm:}A_2{\rm:}Q)$ vanishes if the quantum system $Q$ is
initially in a pure state and if the measurement process is unitary.
But, the ancilla is ``self-consistent'' that is $S(A_2|A_1)=S(A_1|A_2)=0$}.
Clearly, as far as information extraction is concerned,
a more interesting case to consider is the measurement
of a quantum system $Q$ initially prepared in a {\it mixed} state; only then
can information be extracted about the {\it preparation}
of the state. 

A measurement performed on a quantum system
which can be prepared in different states yields an amount
of information about the preparation which is limited by the
Kholevo bound~\cite{bib_kholevo}. 
More precisely, if a system is prepared in a state described
by one of the density operators $\rho_i$ ($i=1,\cdots n$), with 
probability $p_i$, then the information $I$
that can be gathered about the identity of the
state is always lower than the Kholevo bound
\begin{equation}   \label{eq_originalkholevo}
I \le S(\sum_i p_i \rho_i) - \sum_i p_i S(\rho_i)   \;.
\end{equation}
This result holds for any measurement one can perform on the system,
including positive-operator-valued measures (POVM's). Since the original
conjecture by Kholevo, a lot of effort has been devoted to obtaining a more
rigorous proof of the theorem, or to derivations of stronger upper
or lower bounds on $I$
\cite{bib_levitin,bib_schum,bib_yuen,bib_jrw,bib_caves,bib_schumacher}.
Our aim here
is to give a simple proof of this upper bound on accessible information
which is based on quantum entropies, as opposed to
deriving Shannon entropies
from the quantum probabilities associated with measurements,
as is usually done. The derivation relies only on the
unitarity of the measurement seen as a physical process, along with
the strong subadditivity property of quantum entropies ({\it cf.} Sect.~3). 
This makes the physical content of the Kholevo theorem more
transparent: in short, it states that the {\it classical} mutual
entropy (i.e. the acquired information $I$) is bounded from above by
a {\it quantum} mutual entropy.

Let us assume that we have a ``preparer'', described by a
(discrete) internal variable $X$, which is distributed according
to the probability distribution $p_i$ ($i=1,\cdots N$).
The internal state of the preparer, considered as a physical quantum system,
is given by the density matrix\footnote{Of course, $\rho_X$ can be seen
as resulting from the partial trace of a pure state
in an extended Hilbert space (it can be ``purified''
via a Schmidt decomposition).}
\begin{equation}
\rho_X = \sum_i p_i |x_i\rangle \langle x_i|
\end{equation}
with the $|x_i\rangle$ being an orthonormal set of states.
The state of the quantum variable $X$ can be
copied to another system simply by making conditional dynamics (the simplest
example being a controlled-NOT quantum gate)
and in that sense, it behaves like a classical variable
(it can be ``cloned''). Let us therefore denote by $X$ the collective set
of correlated internal variables describing the preparer state.
Assume now that the preparer has at his disposal a set of $N$ mixed states
$\rho_i$, and that he chooses one of them for $Q$ according
to his internal state $X$. The joint state of the preparer and
the quantum system $Q$ is then given by
\begin{equation}
\rho_{XQ} = \sum_i p_i |x_i\rangle \langle x_i| \otimes \rho_i
\end{equation}
and a partial trace over $X$ simply gives the state of $Q$:
\begin{equation}
\rho_Q = {\rm Tr}_X \rho_{XQ} = \sum_i p_i \rho_i \equiv \rho   \;.
\end{equation}
The quantum entropy of $X$, $Q$ and the joint system $XQ$
is given by
\begin{eqnarray}
S(X) &=& H[p_i] \;, \nonumber \\
S(Q) &=& S(\rho) \;, \nonumber \\
S(XQ) &=& H[p_i] + \sum_i p_i S(\rho_i)  \;,
\end{eqnarray}
where the last expression results from the fact that $\rho_{XQ}$
is block-diagonal (it is the quantum analogue of the ``grouping theorem''
in Shannon theory \cite{bib_ash}).
\par

Now, the quantum system $Q$ is ``measured'' by interacting unitarily
with an ancilla $A$, according to
\begin{equation}
\rho_{X'Q'A'} = (1_X \otimes U_{QA}) (\rho_{XQ} \otimes |0\rangle \langle 0|) 
(1_X \otimes U_{QA})^{\dagger}
\end{equation}
where $|0\rangle \langle 0|$ denotes an initial reference state of the
ancilla, and $X'$, $Q'$, and $A'$ correspond to the respective
systems {\em after} the unitary evolution $U_{QA}$.
For the moment, let us assume that $U_{QA}$ is arbitrary.
The interesting question will be to determine the mutual quantum entropy
$S(X'{\rm:}A')$ between the physical state of the ancilla $A$
{\it after} measurement and the physical state of the preparer $X$
(which remains unchanged in the measurement). We will show that,
given certain assumptions for $U_{QA}$, $S(X'{\rm:}A')$
represents simply the Shannon mutual entropy between the preparer
and the ancilla, or,
in other words, the information $I$ extracted by the observer
about the preparer state.
\par

\begin{figure}
\caption {Entropy Venn diagram for the correlated system
$XQ$ before measurement.}
\label{fig_access}
\vskip 0.25cm
\centerline{\psfig{figure=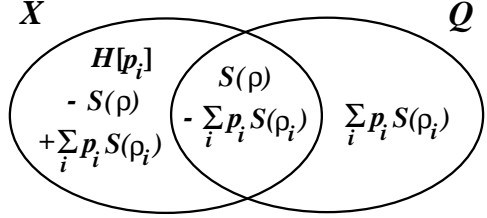,width=2.5in,angle=-90}}
\end{figure}

The relations between the entropies of $X$ and $Q$ 
before measurement can be summarized
by the quantum entropy diagram in Fig.~\ref{fig_access}.
It is easy to calculate the quantum mutual entropy (or mutual
entanglement) between $X$ and $Q$ before measurement,
\begin{equation}
S(X{\rm:}Q)= S(X)+S(Q)-S(XQ)
= S(\rho)- \sum_i p_i S(\rho_i)
\end{equation}
showing that $S(X{\rm:}Q)$ is simply 
the Kholevo bound [see Eq.~(\ref{eq_originalkholevo})].
Invoking the upper and lower bounds for the entropy of a convex combination
of density matrices (see {\it e.g.} \cite{bib_wehrl}), {\it i.e.},
\begin{equation}
\sum_i p_i S(\rho_i) \le S \left(\sum_i p_i \rho_i \right) \le
H[p]+\sum_i p_i S(\rho_i)
\end{equation}
implies
\begin{equation}
0 \le S(X{\rm:}Q) \le H[p_i] \;.
\end{equation}
This shows that the entropy diagram
for $XQ$ (represented in Fig.~\ref{fig_access}) has only positive
entries and therefore looks like a classical diagram for correlated
variables\footnote{As explained earlier, this property is related
to the fact that $\rho_{XQ}$ is a separable state and therefore
is associated with positive conditional entropies.}.
\par

\begin{figure}
\caption {Diagrammatic representation of the Kholevo theorem. The area
enclosed by the double solid lines represents the mutual entropy
that is conserved in the measurement $S(X'{\rm:}Q'A')=S(X{\rm:}Q)$.}
\label{fig_arithmetic}
\vskip 0.25cm
\centerline{\psfig{figure=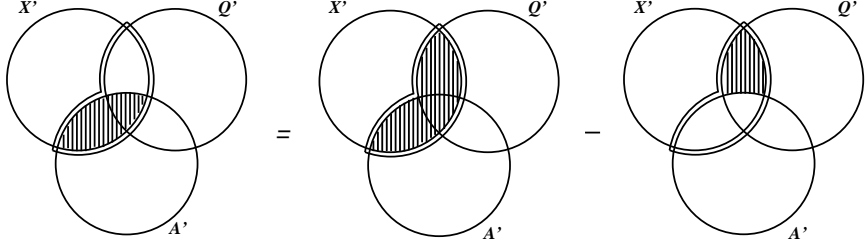,width=4.5in,angle=-90}}
\end{figure}

Before measurement, the ancilla $A$ is in a pure state $|0\rangle$
and the joint state of the system $XQA$ is a product
state $\rho_{XQ} \otimes |0\rangle \langle 0|$, so that we have
$S(X{\rm:}Q) = S(X{\rm:}QA)$. As the measurement involves unitary evolution
of $QA$ and leaves $X$ unchanged, it is straightforward to check
that this mutual entropy is conserved:
\begin{equation} \label{equs1}
S(X'{\rm:}Q'A')=S(X{\rm:}QA)=S(X{\rm:}Q)    \;.
\end{equation}
Next, we may split this entropy according
to the quantum analogue of the chain rules
for mutual entropies [Eq.~(\ref{eq_chainrule2})] to obtain
\begin{equation} \label{equs2}
S(X'{\rm:}Q'A')=S(X'{\rm:}A') + S(X'{\rm:}Q'|A')
\end{equation}
where the second term on the right-hand side is a quantum conditional mutual
entropy ({\it i.e.},
the mutual entropy between $X'$ and $Q'$, conditionally on $A'$).
Combining Eqs. (\ref{equs1}) and (\ref{equs2}) gives the basic relation
\begin{equation}  \label{eq_egalite}
S(X'{\rm:}A')= S(X{\rm:}Q) - S(X'{\rm:}Q'|A')   \;.
\end{equation}
This equation is represented as arithmetic on Venn diagrams
in Fig.~\ref{fig_arithmetic}.

Thus, the quantum mutual entropy between the 
state of the preparer $X'$ (we can ignore the prime since $X$
is unchanged in the measurement) and the state of the ancilla 
after measurement $A'$
is given by $S(X{\rm:}Q)$, the Kholevo bound, reduced
by an amount which represents the mutual entropy still existing between
the preparer's internal variable $X$ and the quantum state
after measurement $Q'$, {\it conditional} on the observed
state of the ancilla $A'$. 
Since $S(X'{\rm:}Q'|A')$ is in general difficult to calculate,
we can make use of strong subadditivity\footnote{Expressed
in our quantum information-theoretic language,
strong subadditivity implies that the conditional mutual entropy
$S(X{\rm:}Y|Z)$ between any three quantum
variables $X$, $Y$, and $Z$ is non-negative. This expresses the intuitive
idea that the mutual entanglement between $X$ and $YZ$ is larger or equal
to the mutual entanglement between $X$ and $Z$ only (just as
mutual informations in Shannon theory), so that a mutual entanglement can
never decrease when extending a system.} in order to obtain an inequality.
In particular, we have $S(X'{\rm:}Q'|A') \ge 0$, which
yields the simple upper bound:
\begin{equation}  \label{eq_kholevo}
S(X'{\rm:}A')\le  S(X{\rm:}Q) = S(\rho) - \sum_i p_i S(\rho_i)   \;.
\end{equation}
It remains to show that, for a particular $U_{QA}$ which
describes a measurement, the quantum mutual entropy
$S(Q'{\rm:}A')$ reduces to a Shannon mutual entropy (the mutual information
$I$ between the state of the preparer and the outcome of the measurement).

Let us consider only the case of a von Neumann measurement\footnote{It
can be shown that the same reasoning applies also to a 
positive-operator-valued measure (POVM) in general.}, 
using the explicit form
\begin{equation}
U_{QA}= \sum_{\alpha} P_{\alpha} \otimes V_{\alpha}
\end{equation}
where the index $\alpha$ refers to the outcome of the measurement and
the $P_{\alpha}$'s denote the projectors in the $Q$ space
associated with the measurement ($\sum_{\alpha} P_{\alpha} =1$).
The unitary operators $V_{\alpha}$ act in the $A$ space, and move
the ancilla from the initial state $|0\rangle$ to a state
$|\alpha\rangle = V_{\alpha} |0\rangle$ that points to the outcome
of the measurement. Let us assume that the $|\alpha\rangle$ are orthogonal
so that the outcomes are perfectly distinguishable. 
The joint density matrix after unitary evolution is thus given by
\begin{equation}
\rho_{X'Q'A'}= \sum_{i,\alpha,\alpha'} p_i |x_i\rangle \langle x_i|
 \otimes P_{\alpha} \rho_i P_{\alpha'}
 \otimes |\alpha\rangle \langle\alpha'|   \;.
\end{equation}
As before, we now have to trace over
the quantum system $Q'$ in order to induce
correlations between $X'$ and $A'$. The corresponding density matrix is
\begin{equation}
\rho_{X'A'}= \sum_{i,\alpha} p_i {\rm Tr} (P_{\alpha} \rho_i)
|x_i\rangle \langle x_i| \otimes |\alpha\rangle \langle\alpha| \;.
\end{equation}
As it is a {\it diagonal} matrix, the relations between the
entropies of $X'$ and $A'$ can be described within Shannon
theory (the quantum definitions of conditional and mutual
entropies reduce to the classical ones in this case.)
A simple calculation shows that one has indeed
\begin{eqnarray}   
S(X'{\rm:}A') &=& H \left[ {\rm Tr}(P_{\alpha}\rho) \right]
  - \sum_i p_i H \left[ {\rm Tr}(P_{\alpha}\rho_i) \right]  \nonumber \\
 &=& H(A)-H(A|X) = H(A{\rm:}X)  \;,    \label{eq_simple}
\end{eqnarray}
where ${\rm Tr}(P_{\alpha}\rho_i)$ is the conditional probability
$p_{\alpha|i}$ of measuring outcome $\alpha$ on states $\rho_i$,
so that it is justified to identify $S(X'{\rm:}A')$ with the
information $I$ in this case. 
Note that the information gained in the measurement is not described
as a difference between initial and final uncertainty of the observer
(involving a calculation of probabilities as it is usually done), but
rather as a quantum mutual entropy.
As a result of Eq.~(\ref{eq_simple}), we see that Eq.~(\ref{eq_kholevo})
provides an upper bound on the accessible information,
and this completes our derivation of the Kholevo theorem. 
As shown elsewhere, the same reasoning
can be extended to the case of sequential measurements
of a quantum system, using chain rules for quantum entropies,
providing a generalization of the Kholevo theorem.

As a final remark, let us mention that
inequality~(\ref{eq_kholevo}) can be shown to be a
special case of a more general relation. For
an arbitrary density matrix $\rho_{XY}$ describing a bipartite quantum
system whose components interact with ancillae $A$ and $B$ that define
bases $|x\rangle$ and $|y\rangle$ respectively, we have clearly
$S(A'{\rm:}B')=H(X{\rm:}Y)$, where $H(X{\rm:}Y)$ is
the Shannon mutual entropy of the joint
probability $p_{xy}=\langle x,y|\rho_{XY}|x,y\rangle$.  Using
\begin{eqnarray}
S(X{\rm:}Y)&=&S(X'A'{\rm:}Y'B')\nonumber\\
&=&S(A'{\rm:}B')+S(A'{\rm:}Y'|B')+S(X'{\rm:}Y'B'|A')
\end{eqnarray}
and the non-negativity of conditional mutual entropies yields
the general inequality
\be
H(X{\rm:}Y)\leq S(X{\rm:}Y)
\ee
between classical and quantum mutual entropies.

\section{Conclusions}

We have shown that quantum entanglement can be
consistently described using the notion of negative conditional
entropy, an essential feature of a quantum information theory built 
entirely on density matrices. Negative quantum entropy can be traced back
to ``conditional'' density matrices which admit eigenvalues larger
than unity. A straightforward definition of quantum mutual entropy,
or mutual entanglement, can also be obtained using a ``mutual''
density matrix. This quantum matrix-based formalism gives rise to
the violation of well-known bounds in classical information theory.
It treats quantum entanglement and classical correlation on the same footing,
while clarifying in which sense entanglement can induce correlation.
This last point allows for a consistent information-theoretic
description of unitary quantum measurement, devoid of any assumption
of wave-function collapse, which, at the same time, accounts for 
the creation of entropy (random numbers) in the measurement outcome.
This sheds new light for example on information-theoretic aspects
of Bell-type experiments or on the issue of how much information
can be accessed in a quantum measurement.
Also, as quantum entanglement is a central feature of quantum
computation, we believe that the present formalism will shed new
light on decoherence (entanglement with an environment) 
in noisy quantum channels, 
as well as the error-correcting codes being devised to counteract it. 
From a more fundamental point of view, the fact that
quantum conditional entropies can be negative
reveals that quantum statistical mechanics is qualitatively very different
from classical statistical mechanics, even though most of the formulae
are similar.

\ack
We would like to thank Hans Bethe, Steve Koonin, and Asher Peres
for very useful discussions.
This work was supported in part by the National Science Foundation  
Grant PHY94-12818 and PHY94-20470, and by a grant from DARPA/ARO
through the QUIC Program (\#DAAH04-96-1-3086).

\end{document}